\documentstyle[twocolumn,prb,aps]{revtex}

\begin{document}

\title{Magnetic properties of {\it 3d}-impurities substituted in GaAs
} 

\author{S. Mirbt, B. Sanyal and P. Mohn$^{*}$ } 

\address{Department of Physics, Uppsala University, 
 Uppsala, Sweden} 
\address{* {\it permanent address} CMS, TU Wien, Wien, Austria} 

\date{15 October 2001}

\maketitle
\begin{abstract}
We have calculated the magnetic properties of substituted 
{\it 3d}-impurities (Cr-
Ni) in a GaAs host by means of first principles electronic structure
calculations. 
We provide a novel model explaining the ferromagnetic long rang order of
III-V dilute magnetic semiconductors.
The origin of the ferromagnetism is shown to be due to delocalized
spin-uncompensated As dangling bond electrons.
Besides the quantitative prediction of the magnetic moments, our model 
provides an understanding of the halfmetallicity,
and the raise of the critical temperature with the impurity
concentration. 

\end{abstract}
\vspace{20mm}
\pacs{PACS numbers: 75.50.Dd, 71.23.-k, 71.55.Eq, 75.30.-m}

\narrowtext
As the electronic device sizes are shrinking, physicists are asked to provide new device concepts.
Devices that work, not despite quantum mechanics but because of quantum mechanics. One track 
is the area of magnetoelectronics, where one in addition to the charge of
the electron tries to exploit
its spin degree of freedom \cite{general}. A central problem is here the spin injection into the semiconductor
\cite{schmidt}. The idea of using a
ferromagnetic semiconductor, i.e. a diluted magnetic semiconductor
(DMS), for the spin injection is now being worked on worldwide \cite{nature}. So far, the
mechanism for the ferromagnetic long range order in DMSs has been a matter of dispute. 
Akai \cite{akai} and more recently Dietl et al. \cite{dietl} proposed the double-exchange mechanism to explain the ferromagnetic order. 
 
In this letter, we propose a novel mechanism for the ferromagnetism of
DMSs. 
We show that the ferromagnetic order is caused by itinerant 
spin-uncompensated As dangling bond (DB) states: 
The {\it 3d}-impurity atom interacts with the As DBs. 
Because of the spin-polarization of the {\it 3d}-atom, the hybridization with
the spin-up As DBs is larger than with the spin-down ones. 
This has two consequences: i) Since
the DB states are delocalized, the hybridization leads to an
itinerant magnetic moment. ii) The filling of the As DBs remains
incomplete. An eventual filling of these holes destroys the
ferromagnetic order in agreement with experiment \cite{ohno2}.  

To support our analysis we perform {\it ab initio} calculations of 
Ga-substituted impurities (Cr,
Mn, Fe, Co, and Ni) in a
GaAs host. The magnetic properties were calculated using a plane wave
pseudopotential code (VASP \cite{VASP1,VASP2}) within the
local spin density approximation (LSDA) \cite{LSDA}. 
The atoms are described by ultra soft Vanderbilt type pseudopotentials
\cite{vanderbildt} as supplied by G. Kresse and J. Hafner \cite{hafner}. The wave functions are expanded in plane waves with an 
energy cut-off of 295 eV.
The electron density was calculated using special {\bf k}-point sets
 \cite{monkhorst} corresponding to a 2\ x\ 2\ x\ 2 folding. The
resulting number of {\bf k}-points was shown to be sufficient for the
here studied properties.
The calculations are performed within a 64 atom supercell, where the {\it 3d}-atom 
is put on a Ga-site in the center of the cell. 

In Fig.1 we show the spin-resolved {\it d}-projected density of 
states (DOS) at the impurity site. For the unrelaxed case (left column), 
i.e. all atoms are
sitting on ideal high symmetry positions, GaAs containing
Cr, Mn, or Co (Fe or Ni) impurities has become a half-metal (metal). 
In the right
column of Fig.1 we show the results for the position, volume, and shape relaxed
case. For Cr or Mn impurities in GaAs, the effect of the relaxation is
negligible, whereas for Fe, Co, and Ni the relaxation almost suppresses the
ferromagnetism. 
Co or Ni impurities in GaAs are non-magnetic, 
for Fe in GaAs the magnetic moment has drastically decreased (Fig.2). 

In order to understand these {\it ab initio} results, we apply an
"impurity-molecule model" \cite{coulson,picoli} which we develop in three steps. i) We consider 
the ideal, i.e unrelaxed, GaAs host with a Ga vacancy. 
At the vacancy site the four surrounding As
neighbours contribute five electrons to the four DBs.
We calculate the Ga-vacancy to be non-magnetic.
The site-projected vacancy DOS is shown in Fig.3. Close to the Fermi
energy ($E_{F}$) a split peak appears in the DOS of the
Ga-vacancy (bottom panel of Fig.3). The two peaks have the following
origin: Three As DBs give rise to the peak at $E_{F}$ (three
top panels), whereas
one As DB gives rise to the peak below $E_{F}$ (panel four of
Fig.3). 
We conclude, that the occupation of the As DBs is as follows: One DB is
completely occupied (peak below $E_{F}$) and three DBs are only
half-occupied (peak at $E_{F}$).

In total the vacancy has 3 holes, i.e. there are 
1.5 holes per spin. In a tetrahedral crystal field the {\it $sp^{3}$}
orbitals split into "{\it s}\ "-like {\it $a_{1}$} orbitals and "{\it
p}\ "-like {\it $t_{2}$} orbitals \cite{picoli}.  
At the vacancy site the {\it $a_{1}$}
state is completely filled and the {\it $t_{2}$} states have 3 holes. 
The {\it $t_{2}$} state of the Ga vacancy is located $0.06 eV $ above
the valence band edge and is thus almost degenerate with the continuum
of extended states \cite{bachelet}. The Ga vacancy {\it $t_{2}$}
state can thus be
treated as a delocalized state. Only then the nonmagnetic state of the
vacancy is understandable.

ii) Next the vacancy is filled with the {\it 3d}-impurity. 
In a tetrahedral crystal field the {\it d}-states are split into
{\it $e$} and {\it $t_{2}$} states, where the {\it $e$} states
lie lower in energy than the {\it $t_{2}$} states \cite{ballhausen}.  
For all {\it 3d} impurities (with the exception of Ni) only the 
spin-up {\it $t_{2}$} ({\it $t_{2}^{\uparrow}$}) states are occupied, i.e.  
the {\it $t_{2}^{\downarrow}$} states are completely empty. 
We assume that the impurity retains its atomic character, i.e. the
occupation of the {\it e} and {\it $t_{2}$} levels follows Hund's rules.

iii) Finally, we consider the interaction of the {\it 3d}-impurity with the four
As DBs. 
It is well known that a substitutional impurity in a semiconductor is
ionized so that the semiconductor holes become filled. One would thus
expect the {\it 3d}-atom to contribute {\it 3} electrons to the
hole-filling, i.e a {\it 3+}-state of the {\it 3d} impurity.
However, for a spin-polarized impurity there are two possibilities:
(1) If the exchange energy is larger than the energy gain (bonding
energy) due to the complete hole filling of the As DBs, only the
spin-up channel will interact with the As DBs, i.e a spin-flip on the
{\it 3d} impurity site is prohibited. (2) If, in contrast to the first
possibility, the bonding energy exceeds the exchange energy, 
{\it 3} electrons will be transferred to the As DBs, i.e a spin-flip on
the {\it 3d} impurity site is allowed. In the following, we will discuss
the two cases separately. 

First assume case (1), i.e that the exchange
energy exceeds the bonding energy. This case agrees with the results for
the unrelaxed first-principles calculation.
Due to their like symmetry and large overlap, only the 
{\it $t_{2}^{\uparrow}$} states of the impurity and the As DB hybridize.
Therefore the 3d-impurity contributes only 1.5 spin-up electrons to
the hole-filling of the As DBs. Because the {\it 3d} impurity states
are localized, not 1.5 but an integer number of two electrons are promoted to the Fermi energy, of which
0.5 electrons have a high probability to be found on the impurity site
itself. Thus one electron is shared between the impurity and the
delocalized As DBs.
The occupation of the impurity atom {\it $t_{2}^{\uparrow}$} level 
is therefore reduced by two electrons (states), i.e. the {\it 3d} atom assumes a
{\it 2+} state. In the following we discuss this charge transfer 
in more detail: 

The As DBs receive 1.5 {\it $t_{2}^{\uparrow}$} electrons. In order to
minimize the energy (especially the kinetic energy of the host), it
seems at first sight, that the optimal situation would be to equally
distribute the 1.5 electrons on both spin-channels of the host, i.e. 0.75 electrons
had to spin flip. But, this apparently optimal situation is not allowed, because of the
following: On the impurity site only spin-up states are available at the
Fermi energy ( due to the exchange splitting of the d-states). Since,
as mentioned above, one promoted electron is shared between the impurity atom
and the delocalized As DBs, the spin of one of the two promoted
electrons is restricted to be spin-up. In order to minimize
the energy, only one electron of the 1.5 transferred {\it $t_{2}^{\uparrow}$} 
electrons, is allowed to be equally distributed among both spin-channels, i.e. only
0.5 electrons are allowed to spin flip. In total then, the spin-up
(spin-down)
channel of the host has received 1.0 (0.5) electron.

An alternative discussion of the charge transfer is the following: Let
us start from the vacancy dangling bonds. As mentioned earlier, three
orbitals of the four delocalized As DBs, are half occupied (0.5 electron per
spin) and one orbital is completely occupied (1 electron per spin)
, because five electrons have to be redistributed on four DBs). 
Assume now that we add one electron to the vacancy and distribute it
equally among both spin channels. 
Adding 0.5 electrons to each spin channel of the As DBs, 
gives rise to the following occupation of the As DBs:
three orbitals are completely occupied ( 1 electron per spin) and one
orbital is completely empty ( 0 electron per spin), because
six (five plus one) electrons have to be redistributed on four DBs. This is an
energetically very favorable situation, because the DOS at
the Fermi energy is drastically reduced to zero, i.e. a bandgap exists.
The 3d-impurity donates to the As DBs 1.5 {\it $t_{2}^{\uparrow}$} electrons,
of which one is used to create the bandgap as described above.  
Now, because not only one electron, but 1.5 {\it $t_{2}^{\uparrow}$} electrons  
are transferred form the impurity to the As DBs, the spin-up channel of
the As DBs receives an additional 0.5 electron. Therefore the spin-down
channel has a bandgap, whereas the spin-up channel has one partly
occupied orbital, i.e a state at the Fermi energy (Fig.1). This
additional 0.5 electron corresponds to one half of the shared
electron. 
This discussion also shows, that any other distribution of the 1.5 {\it
$t_{2}^{\uparrow}$} electrons within the As DBs, will increase the total
energy, because the bandgap in the spin-down channel would have to
disappear. 

In total the impurity has lost 1.5 spin-up electrons and the As DBs have
gained 0.5 spin-down electrons and 1.0 spin-up electrons.
Hence the occupation of the spin-up As DBs exceeds 
the occupation of the spin-down As DBs by 0.5 electrons and this
spin-imbalance is accompanied by an {\it itinerant} magnetic moment, $M_{itinerant}
= 0.5 \mu_{B}$.
Note, that although this amounts to a relatively small moment per formula 
unit, it constitutes the origin of the long
range order, i.e the ferromagnetism of the DMS.
Our model also gives a natural explanation for the observation of so called
"hole mediated ferromagnetism" \cite{ohno2}. 
Namely,  any spin-imbalance (ferromagnetism) will be quenched, 
if the remaining 1.5 holes of the As DBs are filled (disappear) for instance 
by additional donor-dopants.

We can now express the (local) magnetic moment at the impurity site simply as follows:
\begin{equation}
M_{local}=d^{\uparrow} - d^{\downarrow} - N ,
\end{equation}
where N is equal to 1.5 (unrelaxed case), and
$d^{\uparrow},d^{\downarrow}$ correspond to the d-occupation of
the impurity atom.

The total magnetic moment per impurity atom, $M_{total}$, is the sum of the local and itinerant moment:
\begin{equation}
M_{total}=M_{local}+M_{itinerant}=d^{\uparrow}-d^{\downarrow}-1.
\end{equation}
In the case of, for example, (Ga,Mn)As, the local Mn magnetic moment
amounts to $M_{local} = 5 - 0 - 1.5 = 3.5 \mu_{B} $ and $M_{total}$ 
amounts to $4 \mu_B$.
In Fig.2 we summarize our results for Cr, Mn, Fe, Co, and Ni impurities in a GaAs
host. We compare our model (Eq.1-2) with the self consistently calculated magnetic
moments. For the unrelaxed case (left column) discussed so far, 
one notices an overall agreement with some deviations, which however can
be explained easily. 
For Cr and Mn the total magnetic moment agrees perfectly, whereas the
itinerant magnetic moment is smaller by about $0.2 \mu_B $ and the 
local magnetic moment is larger by the
same amount than predicted by our model. In our model, we assumed one
electron to be shared equally between As and the {\it 3d}-impurity. Realizing
that due to the 
difference in electronegativity, more than half of the shared electron is
located on the impurity site, the deviations are obvious and follow the
expected chemical trend.    

The other deviation we find for Ni and Fe. 
This can also be easily understood from the 
calculated {\it d}- and site- projected densities of states (Fig.1). For Fe and Ni we do find
spin-down states at the Fermi energy: in case of Fe these are {\it $e^{\downarrow}$} states 
and in case of Ni {\it $t_{2}^{\downarrow}$} states. 
These additional states contribute to the minimization of the total energy 
and are for simplicity neglected in our model, though it is straight forward
to expand the model. 

Now we compare our model with the calculated DOS (Fig.1).
In agreement with our model,
we find two important features in the spin-up channel: (1) one deep
lying broad feature originating from the localized {
\it $t_{2}$} and {\it $e$}-states on the impurity site ( $E\sim 1.7eV$).
For Cr this broad feature 
does not exist, because the Cr atom only has 2 {\it $t_{2}$} electrons.
Thus for Cr there is an additional empty {\it $t_{2}$} peak above the
Fermi energy.
(2) one double {\it $t_{2}$}-peak close to the Fermi energy originating
from the shared electron between the Ga-vacancy and the impurity.  

So far we have not considered any relaxations that increase the overlap
between the As DBs and the impurity {\it $t_{2}$} states. With an
increased overlap, 
the gain in bonding energy, $E_{bo}$, might become larger than the energy
gain due to the exchange splitting, $E_{x}$. 
Then 3 electrons of the 3d-impurity hybridize with the As DBs, because
the cost to spin-flip 1.5 d-electrons (of the impurity) is compensated 
by the gain in bonding energy. Therefore the impurity will undergo a
transition from a high-spin state to a low-spin state. 
A similar discussion for 3d transition metal ions in Si has earlier
been given by Beeler et al. \cite{beeler}.  
For Cr and Mn the exchange
energy is large and there are no spin-down states at the Fermi level
into which a {\it $t_{2}$} spin up electron could flip. The system therefore
would loose energy when increasing the overlap between As and Cr (Mn). 
That is why the As-Cr and As-Mn bond length is more or less maintained.

For Fe, Co and Ni on the other hand, the system
gains energy, when 1.5 {\it $t_{2}$} electrons spin-flip to fill all 
As DB holes, and consequently the bond length is thereby 
decreased by about 7 \%. (Co,Ga)As and (Ni,Ga)As are non-magnetic 
, whereas $M_{total}$ amounts to $1 \mu_{B}$ for (Fe,Ga)As. 
Our model (N=3 in Eq.1) predicts for Fe a local magnetic
moment of $ 1 \mu_B $, but a zero itinerant magnetic moment. 
With our first principles calculations we find an itinerant moment 
of $0.2 \mu_B$ and a local moment of $0.8 \mu_B$. 
This difference is explained as follows: 
Due to the difference in electronegativity between Fe and As, some of
the delocalized electrons become localized on the Fe site. Since the
{\it $e^{\uparrow}$} states at the Fermi energy are completely
filled, only spin-down electrons can get localized at the
Fe site. It follows, that the local magnetic moment is reduced by the
amount, $A$, of localized spin-down electrons. From Fig.2 
one would expect an $A$ between -0.1 and -0.2 $\mu_{B}$.
This in turn leaves an itinerant magnetic moment of $A\ \mu_{B}$,
in agreement with our first principles
calculations. We thus find the ferromagnetism of (Fe,Ga)As to be
independent on the number of holes in the DMS.  

In order to understand the increase of the critical temperature,
$T_{c}$, with the increase of the Mn impurity concentration
\cite{ohno2}, consider the following:
The itinerant magnetic moment per host atom increases with Mn
concentration, whereas the total magnetic moment per impurity atom is
independent on Mn concentration. Because in our model only the itinerant
magnetic moment is responsible for the long range order, we assume $T_{c}$ 
to be proportional to the itinerant magnetic moment. 
Accordingly, $T_{c}$ increases with Mn concentration. 

Recent theoretical calculations on (Ga,Mn)As report the As nearest
neighbour ({\it nn}) local
magnetic moment to be antiparallel to Mn \cite{shirai,ogawa,sanvito,freeman}. 
But this does not contradict with our model as we will illustrate in the
following: 
First, notice that the spin
down channel of the As DBs is more localized than the spin up channel,
because of the following reason:
The 0.5 additional electrons in the spin-up
channel shifts the spin-up DB states deeper into the valence band, and
hence increases its resonance with the valence band, and in effect the
spin up DB states become more delocalized.

Second, it thus follows that on the As {\it nn} sites, the local As moment 
might be determined by the spin down channel and accordingly the local
As moment may be antiparallel to the impurity moment. The sign of the
local As moment depends thus on the chosen local sphere radius. The
smaller the chosen sphere radius, the more negative becomes (also in our
calculations) the local As moment. 
However, most important, these recent calculations confirm the existence
of an integer total magnetic moment and a non-integer local moment at the
impurity site. 

Experimentally, often a magnetic moment of $5.0 \mu_B $ is measured for
(Ga,Mn)As. We did a calculation where we add
one electron to the unit cell which becomes charge compensated 
by a jellium background. In this way we simmulate the electronic effect
of additional dopants in the unit cell, whose defect-level(s) determine the
Fermi level to lie above the pure (Ga,Mn)As Fermi energy. We
find a total moment of $5.0 \mu_B $ and a local moment of 
$4.0 \mu_B $. Within our model this is understandable in
the following way: Adding one electron, the vaccancy has instead of 3
only 2 holes, i.e. one hole per spin. The energetically most stable
situation is with 3 fully occupied orbitals and one empty orbital.  
Out of the 5 spin-up d-electrons now only one electron is transferred to
the As DBs. This electron will maintain the spin, because the system
would loose energy, when both (spin-up and spin-down) empty orbitals 
would become filled by half an electron. Moreover, this transferred
spin-up electron is delocalized, i.e the itinerant moment is $1.0 \mu_B
$. Therefore, the Curie temperature of the
negatively charged Mn impurity should be higher than of the neutral Mn
impurity.

In summary, the present model explains the ferromagnetic long range order as being caused by itinerant
spin uncompensated {\it $t_2$} orbitals. The simultaneous occurrence of
an itinerant and local magnetic moment is explained. Our model quite
naturally explains the halfmetallicity of (Ga,Mn)As. Moreover, because
the itinerant moment per host atom increases with the impurity concentration,
our model explains the dependence of $T_{c}$ on the impurity
concentration. 

\acknowledgements
We are grateful to the Swedish Foundation for Strategic Research (SSF)
, the Swedish Natural Science Research Council (NFR), and the G\"{o}ran
Gustafsson Foundation for financial 
support. We thank J. Hafner and G. Kresse for giving us the possibility to use
VASP.

\begin{figure}
\caption[fig1]{Calculated {\it d}-density of states projected onto the
{\it 3d}-impurity site. The left (right) column shows the DOS for the unrelaxed (relaxed
) case. The vertical lines indicate the position of the Fermi energy. }
\end{figure}

\begin{figure}
\caption[fig2]{Comparison of the calculated magnetic moments and the
model predictions. The left (right) column shows the unrelaxed (relaxed) case. 
In the first row we show the local magnetic moment, in the second row we
show the itinerant magnetic moment, and in the last row we show the
total magnetic moment.}
\end{figure}

\begin{figure}
\caption[fig3]{Calculated DOS for the unrelaxed vacancy: The
four top panels show the {\it p}-DOS projected onto the four
nearest As neighbours (AsI, AsII, AsIII, AsIV) of the vacancy. The bottom panel
shows the sum of the {\it p}-DOS of the four As
neighbours.}
\end{figure}

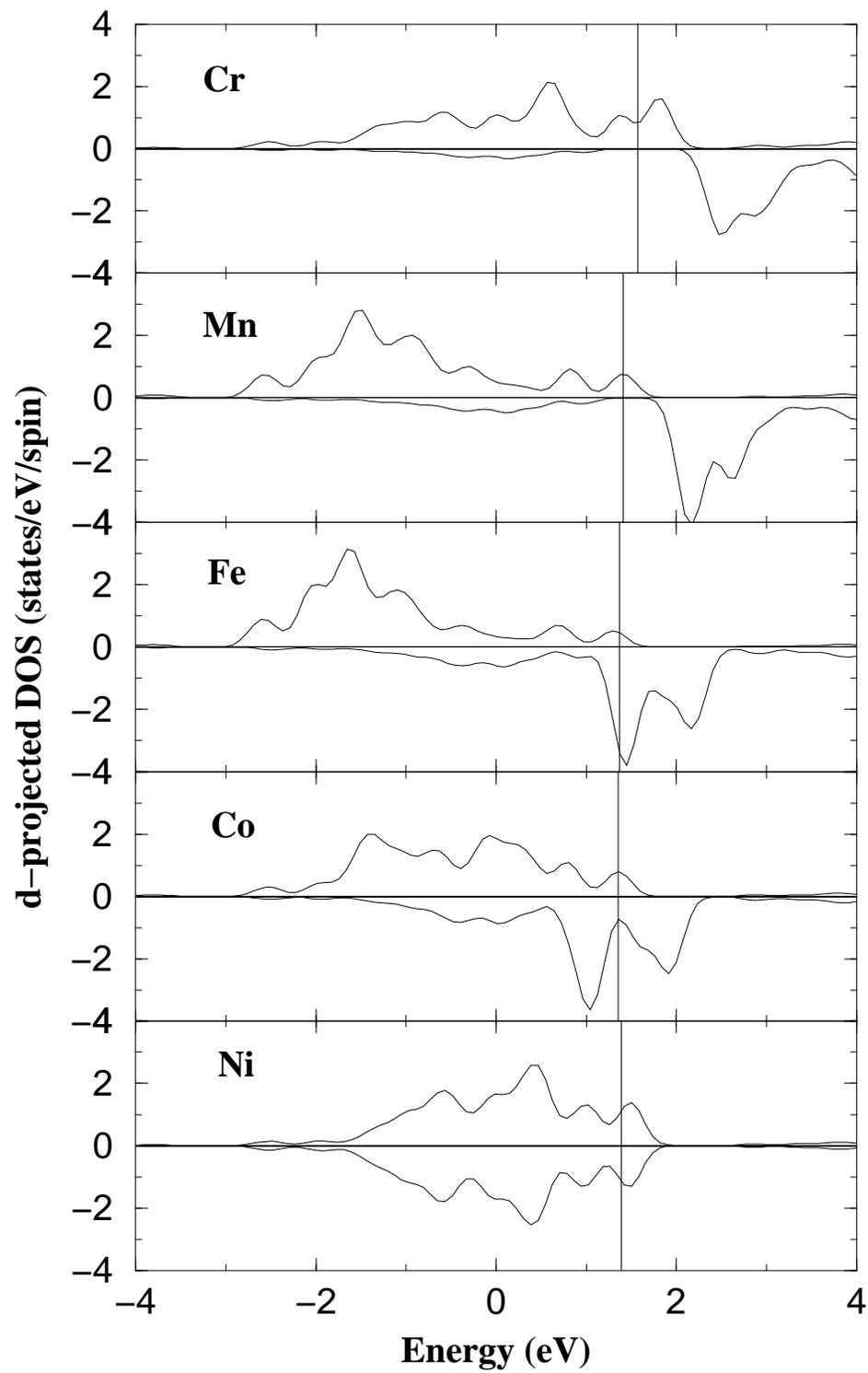
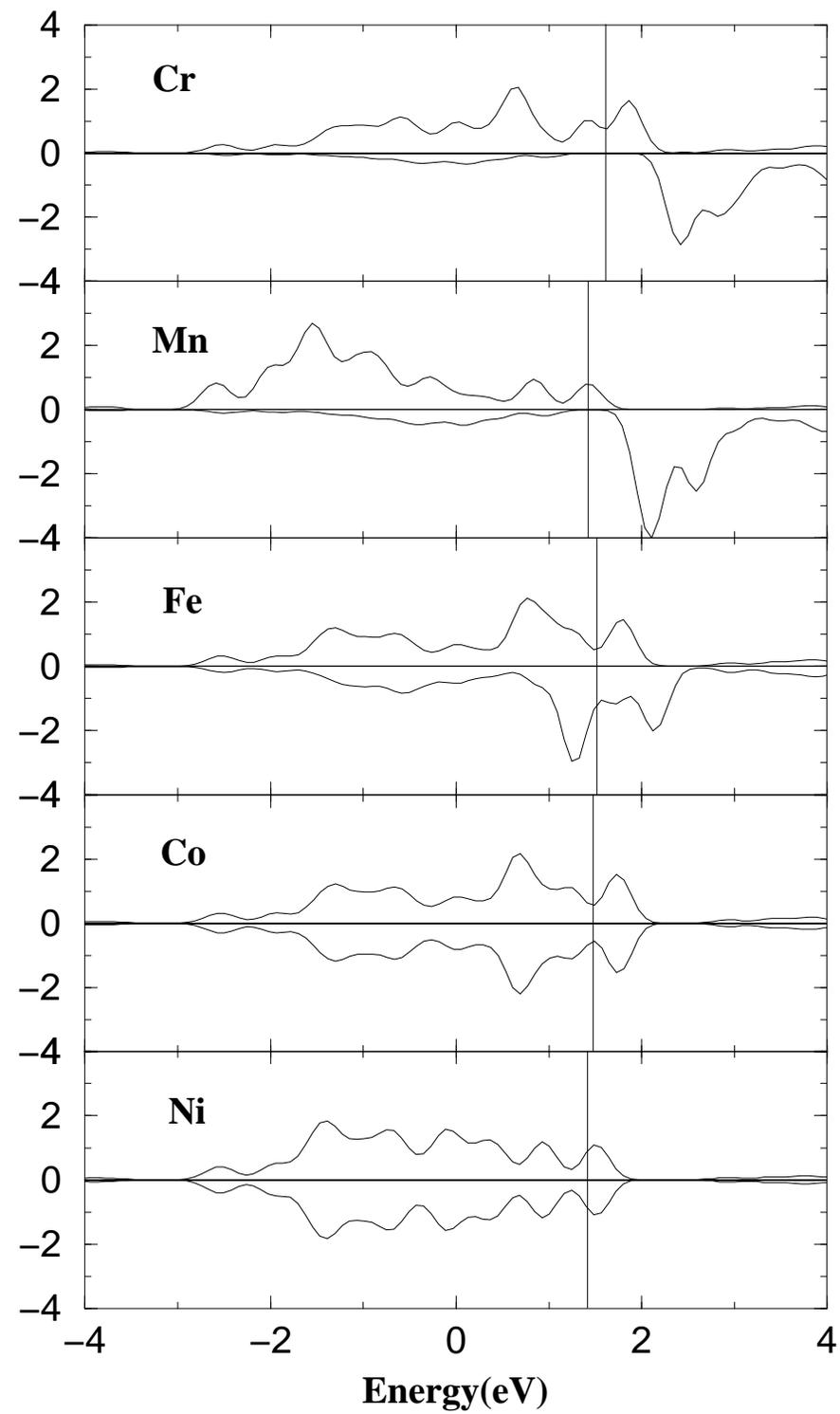
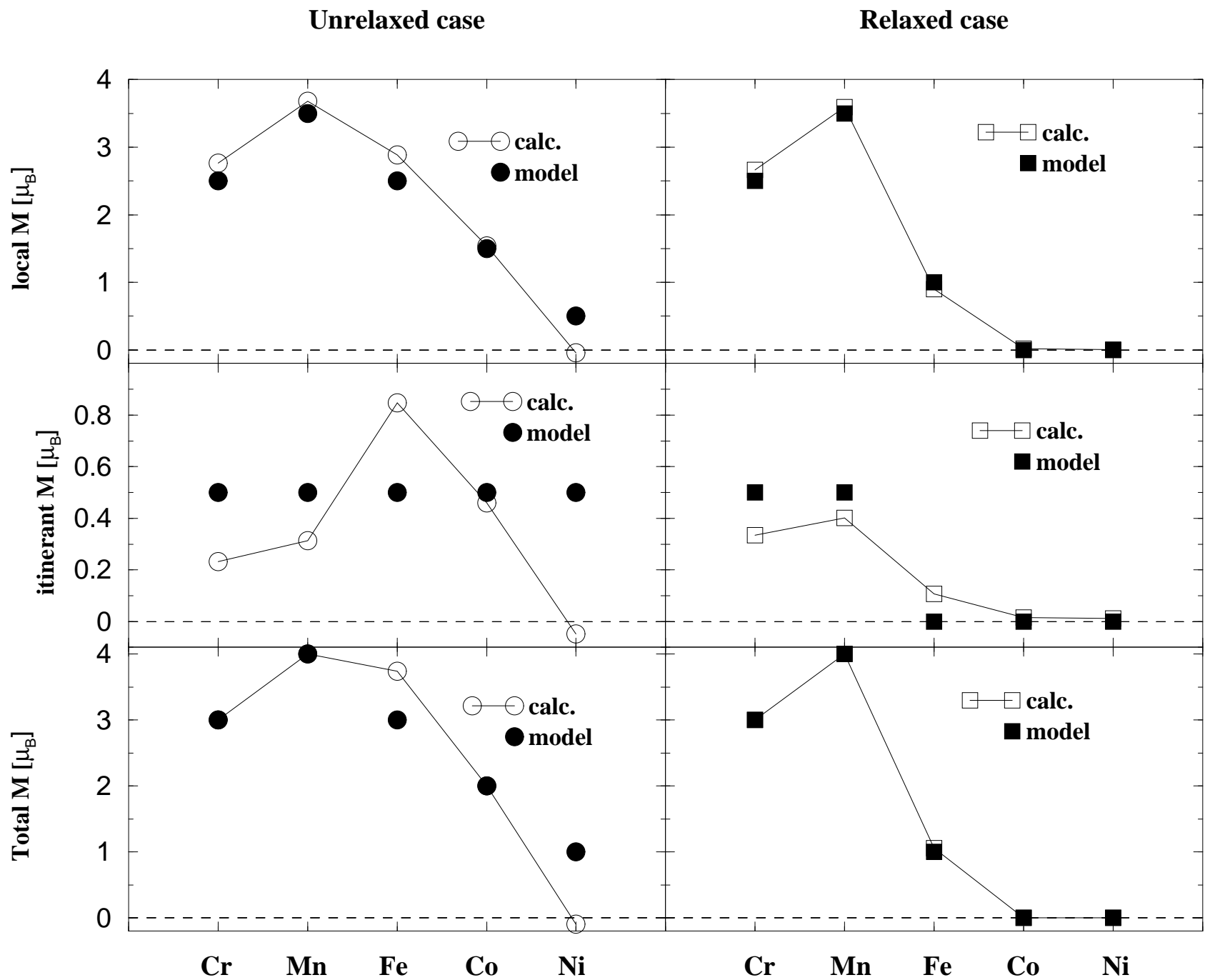
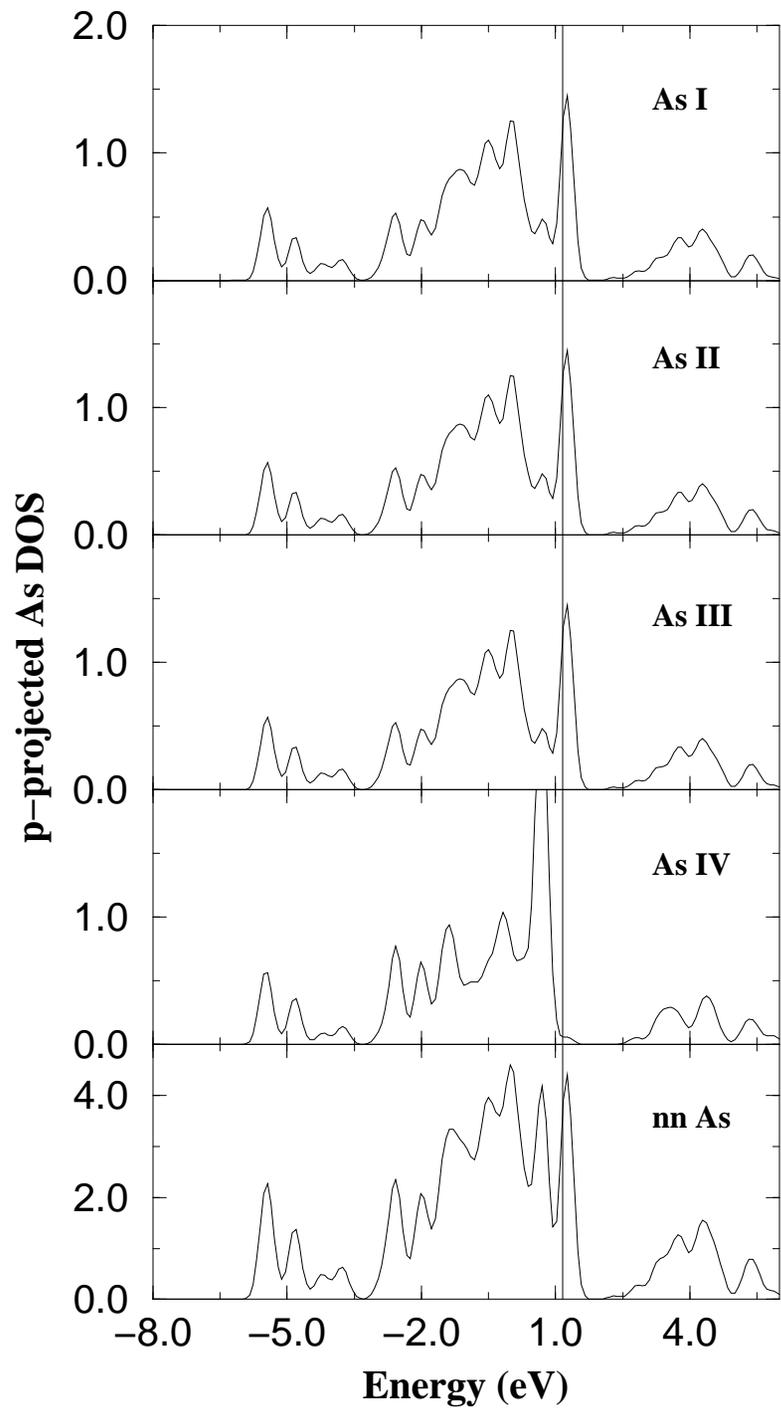
\end{document}